\documentclass[aps,prd,floatfixy,superscriptaddress,showpacs,showkeys]{revtex4}
\usepackage{amssymb}
\usepackage{amsmath}
\usepackage{amsfonts}
\usepackage{epsfig}
\usepackage{graphicx}
\usepackage{graphics}
\usepackage{subfigure}
\usepackage{color}
\usepackage{verbatim}
\newcommand{\eq}{\begin{eqnarray}}
\newcommand{\en}{\end{eqnarray}}

\usepackage[normalem]{ulem}

\renewcommand\sout{\bgroup \color{red} \ULdepth=-.5ex \ULset}
\newcommand{\crct}[2]{\sout{#1} {\color{blue} #2}}
\definecolor{orange}{RGB}{255,127,0}
\definecolor{pink}{RGB}{255,192,203}
\definecolor{brown}{RGB}{139,35,35}
\definecolor{Magenta}{RGB}{255,0,255}

\def\eq{\begin{eqnarray}}
\def\en{\end{eqnarray}}

\def\d12{D_{12}}

\begin{document}

\title{\huge\bf  Selected strong decays of pentaquark State $P_c(4312)$ in a chiral constituent quark model}
\author{Yubing Dong\footnote{dongyb@ihep.ac.cn}}
\affiliation{Institute of High Energy Physics, Chinese Academy of Sciences, Beijing 100049, China}
\affiliation{Theoretical Physics Center for Science Facilities (TPCSF), CAS, Beijing 100049, China}
\affiliation{School of Physical Sciences, University of Chinese Academy of Sciences, Beijing 101408, China}
\author{Pengnian Shen\footnote{shenpn@Ihep.ac.cn}}
\affiliation{College of Physics and Technology, Guangxi Normal University, Guilin  541004, China}
\affiliation{Institute of High Energy Physics, Chinese Academy of Sciences, Beijing 100049, China}
\affiliation{Theoretical Physics Center for Science Facilities (TPCSF), CAS, Beijing 100049, China}
\author{Fei Huang\footnote{huangfei@ucas.ac.cn}}
\affiliation{School of Nuclear Science and Technology, University of Chinese Academy of Sciences, Beijing 101408, China}
\author{Zongye Zhang\footnote{zhangzy@ihep.ac.cn}}
\affiliation{Institute of High Energy Physics, Chinese Academy of Sciences, Beijing 100049, China}
\affiliation{Theoretical Physics Center for Science Facilities (TPCSF), CAS, Beijing 100049, China}
\affiliation{School of Physical Sciences, University of Chinese Academy of Sciences, Beijing 101408, China}
\date{\today}
\begin{abstract}
The newly confirmed pentaquark state $P_c(4312)$ has been treated as
a weakly bound $(\Sigma_c\bar{D})$ state by a well-established
chiral constituent quark model and by a dynamical calculation on
quark degrees of freedom where the quark exchange effect is
accounted for. The obtained mass $4308$~MeV agrees with data. In
this work, the selected strong decays of the $P_c(4312)$ state are
studied with the obtained wave function. It is shown that the width
of the $\Lambda_c\bar{D}^*$ decay is overwhelmed and the branching
ratios of the $p\,\eta_c$ and $p\,J/\psi$ decays are both less than
1 percentage.\\

\end{abstract}
\maketitle
\section{Introduction}
\par\noindent\par

Since the discovery of $X(3872)$ in 2003, the study of XYZ particles has been a hot topic in hadron physics. On the
one hand, many theoretical studies have been carried out in order to understand the peculiar characteristics of these
exotic particles, such as their very narrow widths and the fact that they are very close to the open thresholds.  The
molecular scenario, tetraquark picture, kinematics triangle singularity, and cusp are the most popular interpretations
~\cite{Chen:2016qju,Guo:2017jvc,Esposito:2016noz,Karliner:2017qhf,Dong:2017gaw,Maiani:2004vq,Maiani:2005pe,Maiani:2014aja,
Guo:2019twa,Liu:2019zoy}. On the other hand, many experiments in BEPCII, BELLE, Jefferson Lab., LHCb, etc. have also
been performed for hunting these exotics~\cite{Olsen:2017bmm,Brambilla:2019esw,Ali:2019lzf}.\\

In 2015, in addition to four-quark XYZ meson sectors, LHCb first announced that two pentaquark states $P_c(4310)$
and $P_c(4450)$ were discovered in the decay of $\Lambda_b\to J/\psi pK$ in Run I~\cite{Aaij:2015tga}. Moreover,
they updated their finding in the year of 2019~\cite{Aaij:2019vzc} by careful analyzing the data set including
those collected in Run II. Three, instead of two, pentaquark states were confirmed. Their masses and widths
are~\cite{Aaij:2019vzc}
\begin{eqnarray*}
(M,~\Gamma)_{P_c(4312)}&=&\Big (4311.9\pm 0.7^{+6.8}_{-0.6},~~9.8\pm 2.7^{+3.5}_{-4.5}\Big )~\rm{MeV}, \\ \nonumber
(M,~\Gamma)_{P_c(4440)}&=&\Big (4440.3\pm 1.3^{+4.1}_{-4.7},~~20.6\pm 4.9^{+8.7}_{-10.1}\Big )~\rm{MeV},\\ \nonumber
(M,~\Gamma)_{P_c(4457)}&=&\Big (4457.3\pm 0.6^{+4.1}_{-1.7},~~6.4\pm 2.0^{+5.7}_{-1.9}\Big )~\rm{MeV}.
\end{eqnarray*}
The significance of the new $P_c(4312)$ state is $7.3\sigma$, while such a quantity for $P_c(4440)$ and $P_c(4457)$
is $5.7\sigma$.\\

After $P_c$ states were firstly discovered in 2015, many theoretical investigations have been devoted to the
investigation of the properties of the two exotic baryon states of $P_c(4310)$ and $P_c(4450)$. In particular, the
latest experimental analysis has stimulated a great interest in further understanding of these three pentaquark
states~\cite{Xiao:2019mst,Gutsche:2019mkg,He:2019rva,Wang:2019krd,Lin:2019qiv,Wang:2019spc,Meng:2019ilv,Weng:2019ynv,
Fernandez-Ramirez:2019koa,Voloshin:2019aut,Chen:2019asm}. Various models, such as meson-baryon molecular scenario,
compact five quark states, cusp effect as well as kinematical triangle singularity, etc. have been proposed to
accommodate their structures ~\cite{Chen:2015moa,Chen:2015loa,Liu:2015taa,Takeuchi:2016ejt,Lu:2016nnt,
Yamaguchi:2016ote,Lin:2017mtz}. In fact, even before the LHCb's discovery, theorists have already performed some
model calculations, which can correctly predict the existence of the bound state of the baryon-meson system with
heavy flavor in both quark and hadron degrees of freedom~\cite{Wu:2010jy,Wu:2010vk,Wang:2011rga,Yang:2011wz}. The
chiral constituent quark model is one of them~\cite{Wang:2011rga}.\\

The chiral constituent quark model is a successful model which has been frequently used to calculate the properties
of single hadrons and multi-quark states, especially, the states which may have cluster structures, in a dynamical
and systematical way. As its achievements, the model calculation can well-reproduce the natures of a six light-quark
system with cluster structures, like the phase shifts of the nucleon-nucleon scattering, the binding energy, wave
function with S-D admixture and form factors of the deuteron, the cross sections of the hyperon-nucleon interactions,
and etc., with the same set of model parameters which was fixed before hand by the data of the nucleon-nucleon
scattering and other processes. Meanwhile, it can give a reasonable spectrum of low-lying baryon resonances as well.
More than twenty years ago, this model has been adopted to the study the existence of dibaryon
resonance~\cite{Yuan:1998zys,Shen:1999pf,Li:1999bc,Li:2000cb,Shen:2000cc}, especially the $\Delta\Delta$ resonance
(later called $d^*(2380)$~\cite{Yuan:1999pg}) which eventually was observed by WASA@COSY collaboration in
2014~\cite{Bashkanov:2008ih,Adlarson:2011bh,Adlarson:2012fe,Adlarson:2014pxj} (see the review of
Ref.~\cite{Clement:2016vnl}). Recently, the extracted wave function of $d^*(2380)$ was further employed to calculate the
strong decays of $d^*(2380)$~\cite{Dong:2015cxa,Dong:2016rva,Dong:2017olm,Dong:2017geu,Dong:2018ryf}, the
electromagnetic form factors~\cite{Dong:2017mio,Dong:2018emq}, and the deuteron to $d^*(2380)$ transition form
factors~\cite{Dong:2019gpi}. It is shown that the results with the quark exchange effect on the quark degrees of freedom
fairly-well agree with the experimental data, and therefore one can interpret $d^*(2380)$ as a compact hexaquark
system where the effects of the hidden-color component and the quark exchange play important roles in producing its narrow
widths~\cite{Yuan:1999pg,Brodsky,Clement:2016vnl,Huang:2014kja,Huang:2015nja,Dong:2018ryf}. \\

In analogy to the dibaryon $d^*(2380)$ case, this kind of cluster model calculation was also performed for the
meson-baryon system $(\bar{D}\Sigma_c)$ in 2011~\cite{Wang:2011rga} before the discovery of LHCb. Based on the dynamics
of the chiral constituent quark model, we found a weakly bound state with the mass of $4279-4312$~MeV. Clearly, this
state just corresponds to the observed $P_c(4312)$ resonance with the negative parity. It reflects once more that this
kind of cluster model calculation on the quark degrees of freedom is suitable for unveiling the nature of the
hadron. Especially, when the employed chiral constituent quark model can well-reproduce the existing data, the model
has predictive power. The distinguishing feature of such a treatment, which cannot be involved in the study on the
hadron degrees of freedom, is that the quark exchange effect will be explicitly taken into account in the calculation.
Furthermore, because all the model parameters are pre-determined in previous calculations for the better explanations
of the existing data, no any additional phenomenological form factors or phenomenological vertices among $P_c$,
$\bar{D}$, and $\Sigma_c$ are needed by hand. Therefore, the obtained $P_c$ state is more reliable and less
ambiguity. \\

In this work, we will employe the $P_c(4312)$ wave function obtained
in our previous calculation, where the $P_c(4312)$ state is a weakly
bound $\bar{D}\Sigma_c$ state with a mass of around $4308$~MeV, to
study the  selected strong decays of $P_c(4312)$. Section II gives a
brief introduction of the dynamics of the chiral constituent quark
model. Section III devotes to some selected strong decay processes
of $P_c(4312)$, $P_c(4312)\to J/\psi~p$, $P_c(4312)\to \eta_c~P$ as
well as $P_c(4312)\to \bar{D}^*\Lambda_c$. Finally, we provide
a short summary in section IV.\\

\section{Structure and wave function of $P_c(4312)$ in the extended chiral constituent quark model}
\par\noindent\par

In our previous study of the pentaquark state with heavy flavor, a
system of an open-charm meson and an open-charm baryon is
considered, and a so-called extended chiral SU(3) constituent quark
model (ECCQM) that provides the basic effective quark-quark
interactions caused by the exchanges of the chiral fields, including
pseudo-scalar, scalar and vector mesons, and of one gluon, as well
as the quark confinement, were employed in a dynamical Resonant
group method calculations on the quark degrees of freedom.
Specifically, the Hamiltonian for a 5-quark system is written as \eq
H=\sum_{i=1}^5 T_i-T_G+\sum_{j>i=1}^5 \Big
(V_{ij}^{OGE}+V_{ij}^{Conf.}+V_{ij}^{chv.}\Big ), \en with $T_i$
being the kinetic energy operator for the $i-th$ quark and $T_G$ the
kinetic energy operator for the center of mass (CM)
motion, $V_{ij}^{OGE}$, $V_{ij}^{Conf.}$, and $V_{ij}^{chv.}$
denoting the one-gluon exchange potential, confinement potential,
and the quark-quark potential caused by the chiral field exchanges
between the $i-th $ and $j-th$ quarks, respectively. The latter one
can be abbreviated as
\eq
V_{ij}^{chv.}=\sum_{a=0}^8V_{ij}^{\sigma_a}+\sum_{a=0}^8V_{ij}^{\pi_a}
+\sum_{a=0}^8V_{ij}^{\rho_a},
\en
and the explicit forms of these effective potentials can be found in our previous
papers~\cite{Yuan:1999pg,Huang:2014kja,Huang:2015nja}. In order to
make the model predictive, the determination of parameters must
ensure that as many existing data as possible including the
stability conditions, the masses of the ground state baryons, the
static properties of baryons, the binding energy, root-mean-square
radius (RMS), $S-$ and $D-$ wave admixture in the wave function of
deuteron, the phase shifts of the $N$-$N$ scattering and the cross
sections of the N-hyperon (N-Y) interactions, and even the binding
behavior of $H$-particle and the property of $d^*(2380)$, can all be
well-reproduced. The detailed procedure for the determination of
model parameters can also be
found in our previous papers~\cite{Yuan:1999pg,Huang:2014kja,Huang:2015nja}.\\

The wave function of $P_c(4312)$ obtained in our previous calculation~\cite{Wang:2011rga}
can be  expressed as
\eq
\Psi_{5q,total}^{LSTC}=\sum_ic_i\Psi^{LSTC}(\vec{S_i})
\en
with the superscripts L, S, T, and C denoting the orbital, spin, isospin, and color, and
\eq \Psi^{LSTC}(\vec{S_i})&=&{\cal A}\,\Big [
\phi_A(\vec{\xi}_1,\vec{\xi}_2)\phi_B(\vec{\xi}_3)\chi^L(\vec{r}_{AB}
-\vec{S}_i)\Psi_A^{(STC)_A}\Psi_B^{(STC)_B}~\Big ]\nonumber \\
&=&{\cal A}\,\Big [{\Huge {\Pi}}_{k=1}^3\psi_A\big
(\vec{r}_k,\frac{\mu_{AB}}{M_A}\vec{S}_i\big ) \otimes {\Huge
\Pi}_{l=4}^5\psi_B\big (\vec{r}_l,-\frac{\mu_{AB}}{M_B}\vec{S}_i\big
)\Psi_A^{(STC)_A} \Psi_B^{(STC)_B}~\Big ] , \en where the wave
function of the CM motion is omitted in the first row, because we
are working in the CM system (the unitary transformation between two
sets of coordinates in the first and second rows, respectively, can
be found in Appendix). In the equality above, ${\cal A}$ denotes the
anti-symmetrization operator which can be written as
\eq\label{eq:antisym} {\cal A}=\big
(1-\hat{P}^{OSFC}_{14}-\hat{P}_{24}^{OSFC}-\hat{P}_{34}^{OSFC}\big
), \en where $\hat{P}_{ij}^{OSFC}$ is an exchange operator which
exchanges the $i$-th quark in cluster A and the $j$-th quark in
cluster B in the orbital, spin, flavor and color spaces, and the
first and other three terms on the right hand side of eq.~(\ref{eq:antisym}) are the direct term and
exchange terms in order. $\Psi_{A,B}^{(STC)_{A,B}}$ is the wave
function in the spin-flavor-color space for cluster $A$ or $B$, and
consequently, the color wave function of the direct term of $P_c$ can be expressed as
\eq
\Psi^{C}_{5q}=\frac{1}{\sqrt{6}}\sum_{ijk}\epsilon_{ijk}q^{(1)}_iq_j^{(2)}q_k^{(3)}
\otimes\frac{1}{\sqrt{3}}\delta_{kl}q^{(4)}_k\bar{q}^{(5)}_l,
\en
and the spin-flavor wave function of the  direct term
of $P_c$ can be written as
\eq\label{eq:spinisospin}
\Psi^{(SF)}_{5q}=\frac{1}{\sqrt{2}}\Big
[\chi_{\rho}\xi_{\rho}+\chi_{\lambda}\xi_{\lambda} \Big
]^{(123)}_{(SF)_{\Sigma_c}} \otimes \Big (q^{(4)}\bar{q}^{(5)}\Big
)_{(SF)_{\bar{D}}},
\en
with $\xi_{\rho,\lambda}$ (or
$\chi_{\rho,\lambda}$ ) being the isospin (or spin) wave function
with the mixed-antisymmetry $\rho$ and mixed-symmetry $\lambda$,
respectively, and the number in parentheses in the superscript
indicating the label of the quark. The detailed forms of
$\xi_{\rho,\lambda}$ (or $\chi_{\rho,\lambda}$) can be found in the
standard text book, for instance Ref.~\cite{close}.
$\phi_A(\vec{\xi}_1,\vec{\xi}_2)$ for quarks labeled 1,2,3 and
$\phi_B(\vec{\xi}_3)$ for quarks labeled 4,5 represent,
respectively, the internal wave functions of the baryon and meson
clusters in the coordinate space: \eq
\phi_A(\vec{\xi}_1,\vec{\xi}_1)&=&\Big
(\frac{\omega}{\pi}\frac{m_1m_2}{m_{12}}\Big )^{3/4} \Big
(\frac{\omega}{\pi}\frac{m_{12}m_3}{m_{123}}\Big )^{3/4} \exp\Big
[-\frac{\omega}{2} \Big
(\frac{m_1m_2}{m_{12}}\vec{\xi}_1^2+\frac{m_{12}m_3}{m_{123}}\vec{\xi}_2^2\Big
)\Big ]
\\ \nonumber
&=&\Big (\frac{\omega}{\pi}\mu_{12}\Big )^{3/4}\Big (\frac{\omega}{\pi}\mu_{(12),3}\Big )^{3/4}
\exp\Big [-\frac{\omega}{2}\Big (\mu_{12}\vec{\xi}_1^2+\mu_{(12),3}\vec{\xi}_2^2\Big )\Big ],
\en
with $m_{12}=m_1+m_2$, $\mu_{(12,3)}=\frac{m_{12}m_3}{m_{12}+m_3}$, $\vec{\xi}_1=\vec{r}_2-\vec{r}_1$,
$\vec{\xi}_2=\vec{r}_3-\frac{m_1\vec{r}_1+m_2\vec{r}_2}{m_1+m_2}$, and
\eq
\phi_B(\vec{\xi}_3)&=&\Big (\frac{\omega}{\pi}\frac{m_4m_5}{m_{45}}\Big )^{3/4}
\exp\Big [-\frac{\omega}{2}\frac{m_4m_5}{m_{45}}\vec{\xi}_{3}^2\Big ] =\Big (\frac{\omega}{\pi}\mu_{45}\Big )^{3/4}
\exp\Big [-\frac{\omega}{2}\mu_{45}\vec{\xi}_{3}^2\Big ],
\en
with $\vec{\xi}_3=\vec{r}_5-\vec{r}_4$.
Functions
\eq
\chi^L(\hat{r}_{AB} -\vec{S}_i)&=&\Big (\frac{\omega}{\pi}\mu_{AB}\Big )^{3/4}\exp\Big
[-\frac{\omega}{2}\mu_{AB}\Big (\hat{r}_{AB} -\vec{S}_i\Big )^2\Big ],
\en
with $\mu_{AB}=\frac{(m_1+m_2+m_3)(m_4+m_5)}{m_1+m_2+m_3+m_4+m_5}$, form a set of basis functions peaked locally
at $\vec{S}_i$, which span a domain in the coordinate space for expanding the unknown relative motion wave
function  between two clusters, and $c_i$ are expansion coefficients which are determined by solving the
bound state RGM equation~\cite{Huang:2014kja,Huang:2015nja}:
\begin{equation}\label{eq:RGM-bound}
\langle \delta\Psi_{5q} | H-E | \Psi_{5q} \rangle = 0
\end{equation}
for $P_c$.
\\

In the practical calculation, we only consider the single configuration $\bar{D}\Sigma_c$. This is because that
the effect of the channel mixing from other configurations, like $\bar{D}\Lambda_c$, hidden-color state $|c_8c_8>$,
and etc. is negligible due to the large threshold difference (in fact, our previous calculation confirmed such
a statement).
Now, if we consider the spin-parity of this pentaquark state is $1/2^-$, the resultant mass
of the state is around $4308~\rm{MeV}$, which is very close to the experimental data $4311.9\pm 0.7^{+6.8}_{-0.6}$.
Meanwhile, we also obtain the wave function of this $P_c$ state.\\

\section{Selected strong decays of $P_c(4312)$}

Nowadays, confirming the structure of $P_c$ is an urgent task for the research on the existence of pentaquark states.
One of the most effective ways is to theoretically study its decay modes and corresponding partial decay widths, and
to provide a reference for further verification by experiments. To this end, we select some dominant decay channels,
for instance the $J/\psi+p$, $\eta_c+p$, $\bar{D}^*+\Lambda_c$, $\bar{D}+\Lambda_c$ channels, to study.
Since we assume that the initial state has a $\Sigma_c+\bar{D}$ structure, the decayed particles are
different from the clusters in the initial state. Therefore, there must be quark re-combination or re-arrangement in
the decay processes. In fact, that kind of process has been discussed and considered in the meson-meson scattering
~\cite{Barnes:2000hu,Li:2007sya,Yang:2017nmv}, as well as the $N\bar{N}$ annihilation  processes
~\cite{Maruyama:1980rh,Furui:1984rr,Green:1984ef} in the past. We believe that such effect will also play a role
in the decay of $P_c$. Now, in terms of the obtained wave function of $P_c$ in the bound state calculation,
we include this effect in the $P_c(4312)$ decays. \\

Then, the decay width of the $P_c(4312)$ state to two hadrons, for instance, $J/\psi$ and $p$, in our
non-relativistic chiral constituent quark model, can be written as
\eq \label{eq:width}
\Gamma =\frac{1}{2S_{P_c}+1}\int (2\pi)\delta\Big
(M_{P_c}-E_{J/\psi}(k)-E_p(k)\Big )d^3k \overline{\big |{\cal
M}_{if}(k)\big |^2},
\en
where $\vec{k}$ denotes the three-momentum of the relative momentum between two outgoing hadrons, $c\bar{c}$
(can be either $J/\psi$ or $\eta_c$) and $p$ with $\vec{k}=-\frac{M_{c\bar{c}}}{M_p+M_{c\bar{c}}}\vec{P}_{p}
+\frac{M_{p}}{M_{P}+M_{c\bar{c}}}\vec{P}_{c\bar{c}}$, where $\vec{P}_{c\bar{c}}$ and $\vec{P}_p$ are the
momenta of the outgoing $c\bar{c}$ and proton, respectively.  The magnitude of the relative
momentum $k$ can be expressed as
\eq
k=\lambda^{1/2}\big (M^2_{P_c},
M^2_{J/\psi}, M^2_p\big )\Big /(2M_{P_c}),
\en
with the K\"allen function
\eq
\lambda(x,y,z)=x^2+y^2+z^2-2xy-2xz-2yz.
\en

\subsection {Hidden charm decays}\par\noindent\par

Because the $P_c$ states were discovered in the $\Lambda_b^0\rightarrow J/\psi\,p\,K^-$ process at LHCb, it is
necessary to study the hidden-charm decay of the $P_c$ state. For the decay mode $P_c\to (c\bar{c})p$ (where
$c\bar{c}=J/\psi~\rm{or}~\eta_c$) we consider the re-arrangement of the heavy quark (the $3$-rd quark in the
initial $\Sigma_c$-cluster) and the light quark (the $4$-th quark in the initial $\bar{D}$-cluster) shown in Fig. 1.

\begin{figure}[htbp]
\begin{center}
\includegraphics [width=7.5cm,height=2.5cm]{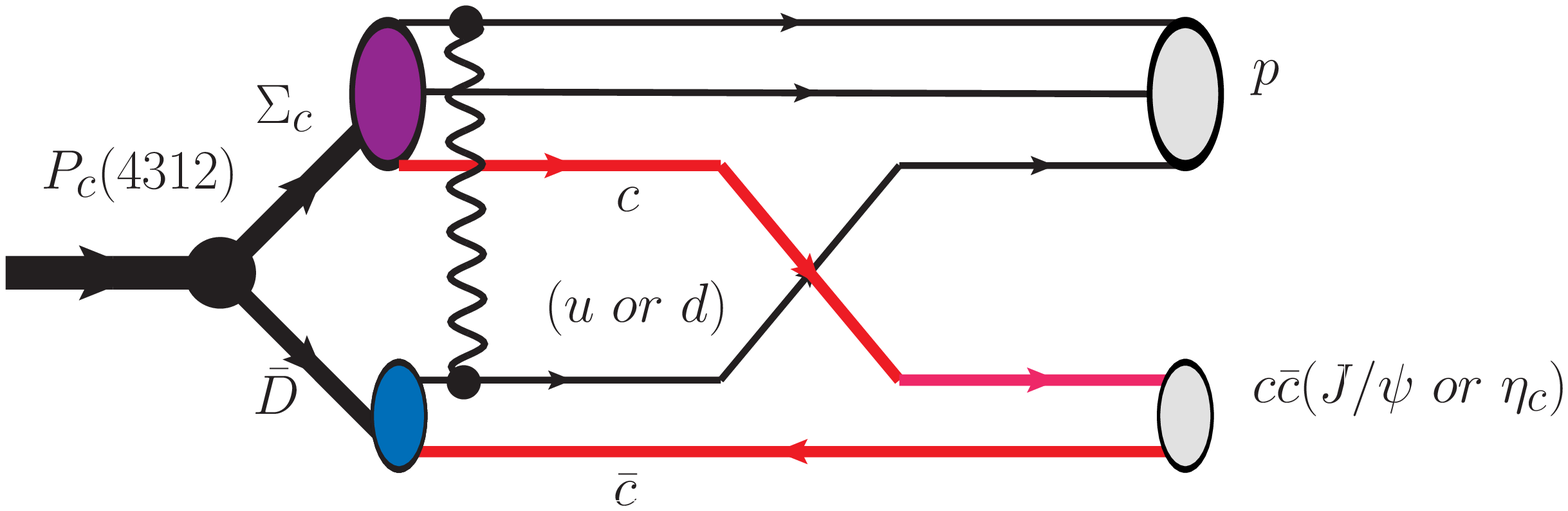}
{\hskip 0.75cm}
\includegraphics [width=7.5cm,height=2.5cm]{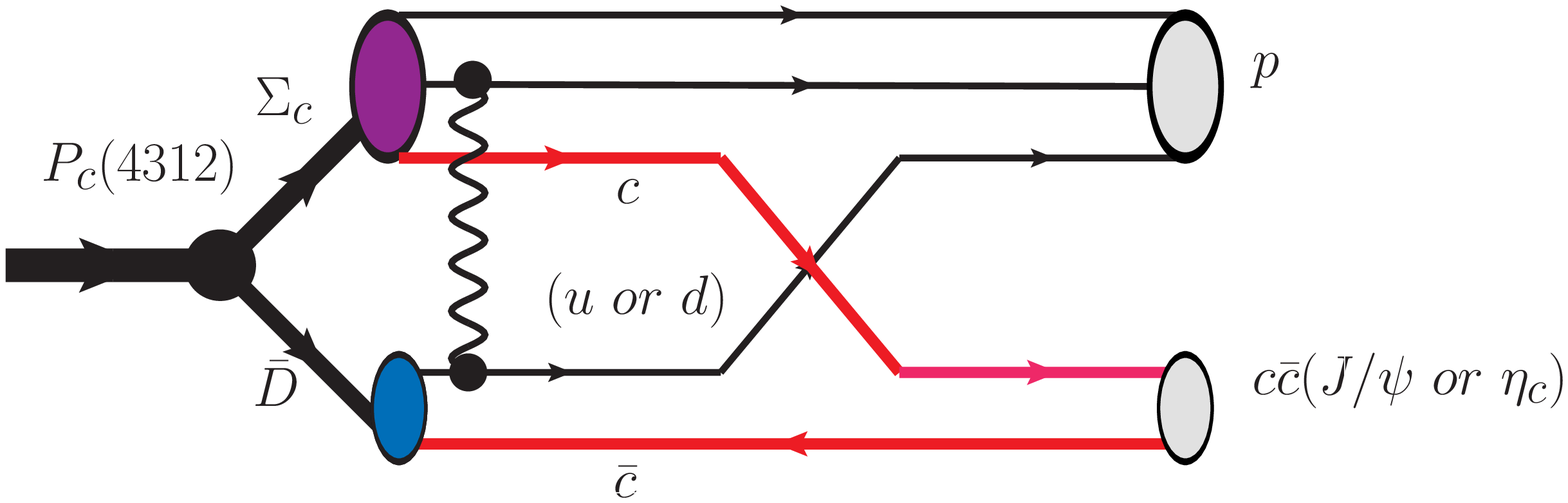}
\caption{An illustration for the hidden charm decays of $P_c(4312)$ with the quark re-arrangement. The wiggle line
stands for the interaction between the two quarks in the meson and baryon clusters, respectively. The light quarks
are represented by the thin black lines and the heavy quarks by the red thick line.}
\label{fig:Fig1}
\end{center}
\end{figure}

Similar to eq.~({\ref{eq:spinisospin}}), the spin-isospin wave function of the final $p$ and $J/\psi$ states is
given by
\eq\label{eq:spinisospinB}
\Psi^{(SF)}_{5q}=\frac{1}{\sqrt{2}}\Big [\chi_{\rho}\xi_{\rho}+\chi_{\lambda}\xi_{\lambda}
\Big ]^{(124)}_{(SF)_{p}} \otimes \Big (q^{(3)}\bar{q}^{(5)}\Big )_{(SF)_{J/\psi}},
\en
where the superscript $(124)$ indicates that quarks in $p$ are labeled by 1, 2, and 4, and $(3)$ and $(5)$
denotes that quarks in $J/\psi$ are marked by 3 and 5, respectively. The detailed forms of $\xi_{\rho,\lambda}$
(or $\chi_{\rho,\lambda}$) can also be found in Ref.~\cite{close}.\\

The orbital wave function of the final state after considering the quark re-arrangement can be expressed as
\eq
\Psi_{f}=\phi_{A'}(\xi_1',\xi_2')\phi_{B'}(\xi_3')\frac{\exp(i\vec{k}\cdot\vec{r}^{~'}_{A'B'})}{\sqrt{V}}
\frac{\exp(i\vec{{\cal P}}\cdot \vec{R}'_{A'B'})}{\sqrt{V}}
\en
with
\eq \phi_{A'}(\vec{\xi}_1',\vec{\xi}_2')&=&\Big
(\frac{1}{b^2_{p_{1}}\pi}\Big )^{3/4} \Big
(\frac{1}{b^2_{p_{2}}\pi}\Big )^{3/4} \exp\Big
[-\frac{1}{2b^2_{p_{1}}}\vec{\xi}_1^{~'2}-\frac{1}{2b^2_{p_{2}}}\vec{\xi}_2^{~'2}
\Big ]\\ \nonumber &=&\Big (\frac{1}{b^2_{p_{1}}\pi}\Big )^{3/4}\Big
(\frac{1}{b^2_{p_{2}}\pi}\Big )^{3/4} \exp\Big
[-\frac{1}{2b_{p_{1}}^2}\vec{\xi}_1^2-\frac{1}{2b_{p_{2}}^2} \big
(\vec{r}_{AB} -\frac{m_c}{m_{123}}\vec{\xi}_2
+\frac{m_c}{m_{45}}\vec{\xi}_3 \big )^2 \Big ], \en and \eq
\phi_{B'}(\vec{\xi}_3')&=&\Big (\frac{1}{b^2_{c\bar{c}}\pi}\Big
)^{3/4} \exp\Big [-\frac{\vec{\xi}_3^{~'2}}{2b^2_{c\bar{c}}}\Big ]
\\ \nonumber
&=&\Big (\frac{1}{b^2_{c\bar{c}}\pi}\Big )^{3/4}\exp\Big [-\frac{1}{2b^2_{c\bar{c}}} \Big
(\vec{r}_{AB}+\frac{2m_l}{m_{123}}\vec{\xi}_2-\frac{m_l}{m_{45}}\vec{\xi}_3
\Big )^2\Big ]
\en
where $b^2_{c\bar{c}}=\frac{2}{\omega m_c}$, $b^2_{p_1}=\frac{2}{\omega m_\ell}$,
and $b^2_{p_2}=\frac{3}{2\omega m_\ell}$  are the harmonic oscillator width parameters of $c\bar{c}$ and proton
systems, respectively, and
\eq
\exp(i\vec{k}\cdot\vec{r}^{~'}_{A'B'})=\exp\Big (\frac{i}{6}\vec{k}\cdot \Big [\vec{r}_{AB}
-2\frac{m_{15}}{m_{123}}\vec{\xi}_2-\frac{m_{15}}{m_{45}}\vec{\xi}_3\Big ]\Big ),
\en
describes the relative wave function between the final $c\bar{c}$ and proton, and $\exp(i\vec{{\cal P}}\cdot
\vec{R'}_{A'B'})$ denotes the wave function of the CM motion with the coordinate of the CM motion
$\vec{R}'_{A'B'}=\vec{R}_{AB}$ and the momentum of the 5 quark system ${\cal P}$ which obeys the
momentum conservation rule $\vec{{\cal P}}=\vec{P}_{c\bar{c}}+\vec{P}_p=\vec{P}_{\Lambda_c}+\vec{P}_{\bar{D}}$.\\

It should be mentioned that as a preliminary study of the decay of $P_c$, this work just plans to give a qualitative
estimate. Therefore, the matrix element $<f|\hat{P}_{34}{\cal H}_T|i>$ will be roughly taken as the product of the
averaged interaction $\overline{{\cal H}}_{Int}$ and the overlap between the initial and final states where
quark re-arrangement effect is involved
\eq &&<f|\hat{P}_{34}{\cal H}_T|i>= \int
d^3r_1d^3r_2d^3r_3d^3r_4d^3r_5 \Bigg
\{\Psi_{c\bar{c}}(\xi_3')\Psi_p(\xi_1',\xi_2')\frac{e^{i\vec{k}\cdot\vec{r}'_{A'B'}}}{\sqrt{V}}
\frac{e^{i\vec{{\cal P}_f}\cdot\vec{R}'_{A'B'}}}{\sqrt{V}}\Bigg \}^*\times\overline{{\cal H}}_{Int}.\nonumber \\
&&~~~~~\times \Bigg \{ \sum_{i=1}^{10}c_i{\cal A}\int
d\hat{S}_i{\Large \Pi}_{k=1}^3
\psi_A(\vec{r}_k,\frac{\mu_{AB}}{M_A}\vec{S}_i) \times
\Pi_{l=4}^5\psi_B(\vec{r}_l,-\frac{\mu_{AB}}{M_B}\vec{S}_i)Y_{LM}(\hat{S}_i){\tilde\Phi}^{CSF}\Bigg
\}.
\en
In order to simplify the calculation without losing the main character, the averaged interaction
$\overline{{\cal H}}_{Int}$ is further approximated by the binding energy of the $P_c(4312)$ state. In our
practical calculation, the experimental values of the masses of mesons and baryons are used, namely
$M_{\Sigma_C^+\bar{D}^0}=4317.73~\rm{MeV}$ and $M_{\Sigma_C^{++}D^-}=4323.62~\rm{MeV}$. Considering the average
value of $M_{\Sigma_C^+\bar{D}^0}$ and $M_{\Sigma_C^{++}D^-}$ and our calculated $P_c$ mass of about
$4308$~MeV, $\overline{{\cal H}}_{Int}$ is taken around $12$~MeV. \\

\subsection {Open charm decay}\par\noindent\par

On the other hand, it is said that the open-charm channel dominates the decay of $P_c$. Therefore, we should also
calculate the partial decay width of such a channel. Here, we consider the quark re-arrangement effect of light
quarks between the initial $\Sigma_c$ and  $\bar{D}$ hadrons. The detailed calculation for this mode is similar
to the hidden charm decay except that the re-arrangement here contains exchange ${\hat P}_{14}$ and exchange
${\hat P}_{24}$, respectively. The schematic diagrams of this decay are shown in Fig. 2.\\

\begin{figure}[htbp]
\begin{center}
\includegraphics [width=7.5cm,height=2.5cm]{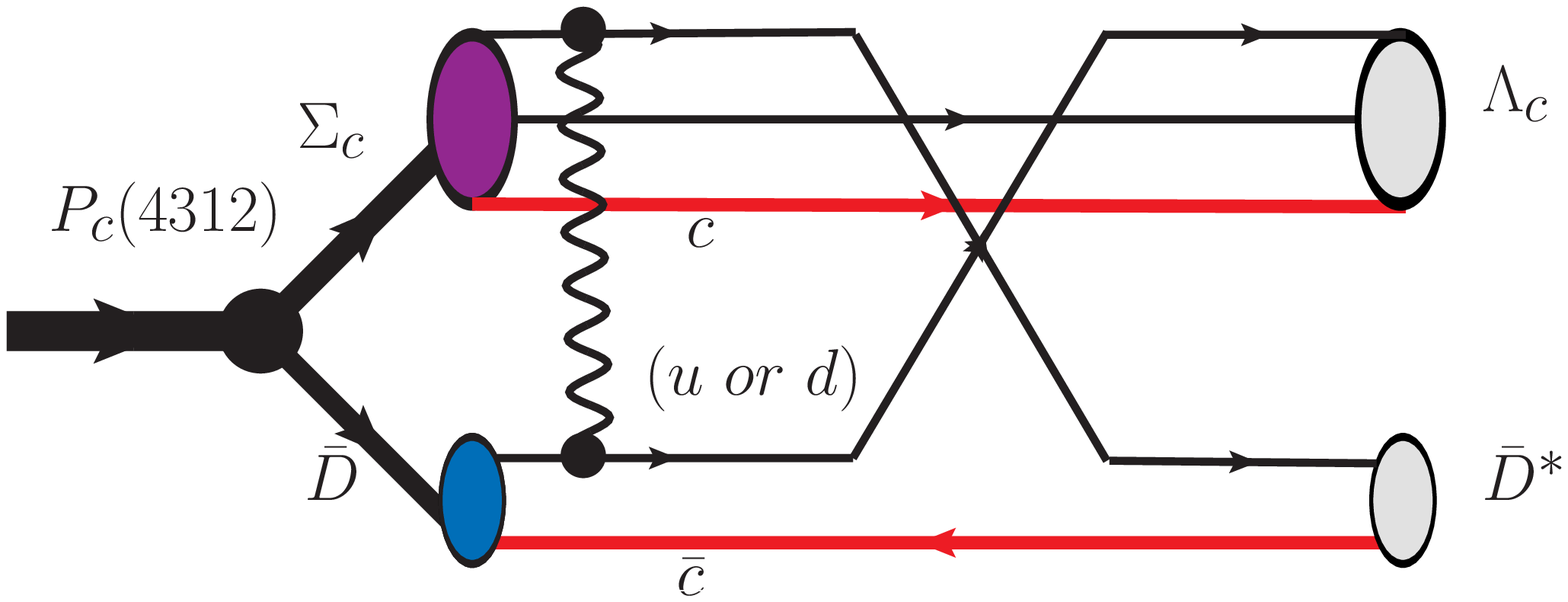}
{\hskip 0.5cm}
\includegraphics [width=7.5cm,height=2.5cm]{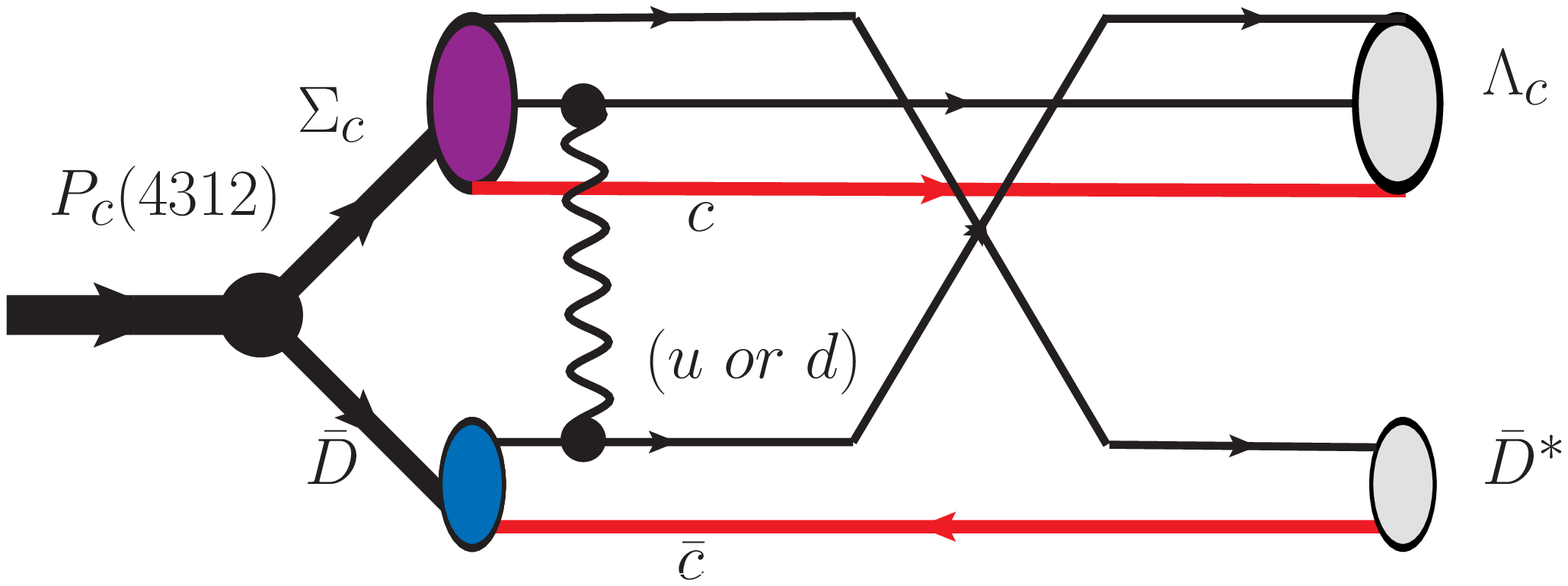}\par

\vspace{1cm}
\includegraphics [width=7.5cm,height=2.5cm]{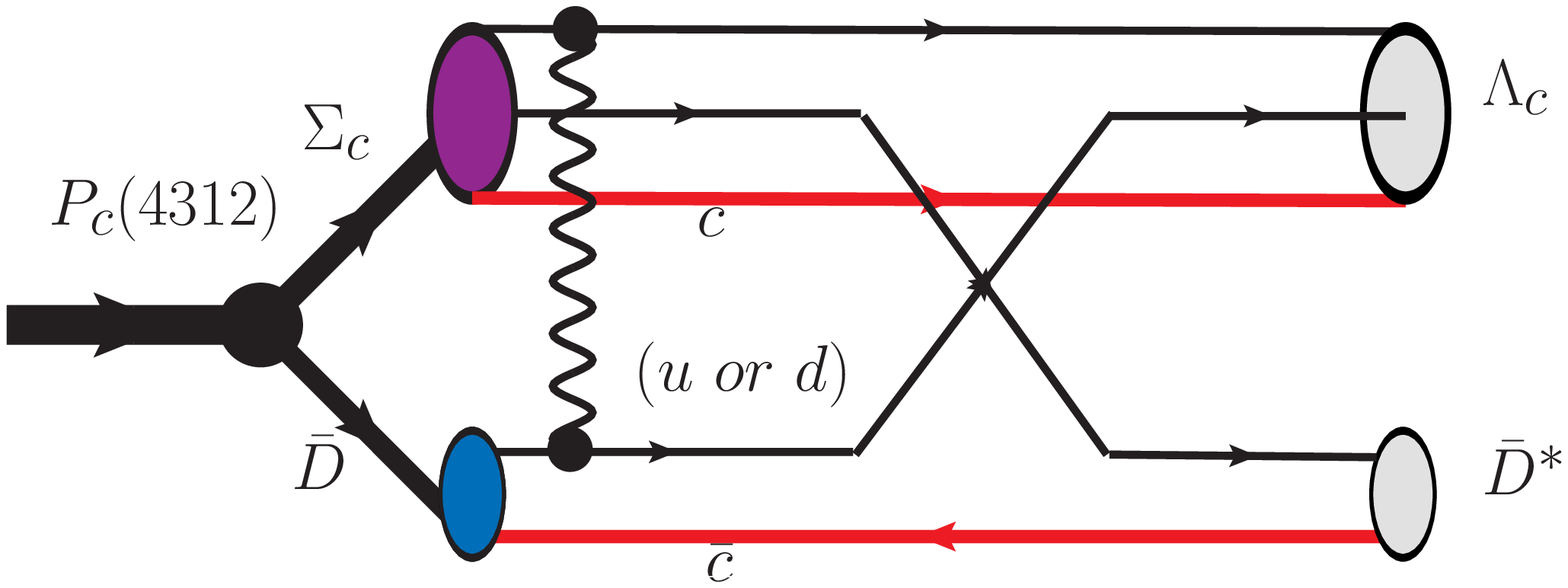}
{\hskip 0.5cm}
\includegraphics [width=7.5cm,height=2.5cm]{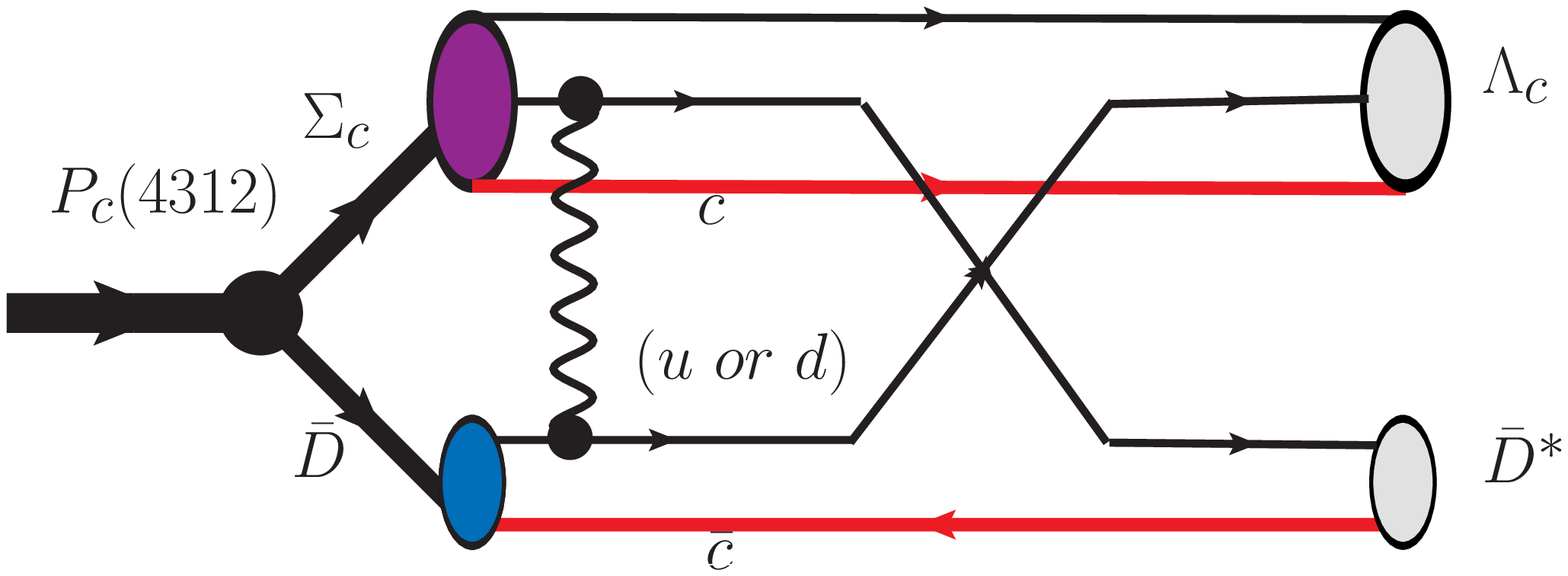}
\caption{An illustration for the $P_c(4312)$ open charm decays with quark re-arrangement. The two figures in the upper
panel stand for the re-arrangement between the 1-st and 4-th quarks, and the two in the lower panel represent the
re-arrangement between the 2-nd and 4-th quarks, respectively.}
\label{fig:Fig2}
\end{center}
\end{figure}

Similar to eq.~({\ref{eq:spinisospin}}), the spin-isospin wave function of the final $\Lambda_c$ and $\bar{D}^{(*)}$
states is given by
\eq\label{eq:spinisospinC}
\Psi^{(SF)}_{5q}=\frac{1}{\sqrt{2}}\Big
[\chi_{\rho}\xi_{\rho}+\chi_{\lambda}\xi_{\lambda}\Big
]^{(423(143))}_{(SF)_{\Lambda_c}} \otimes \Big
(q^{(1(2))}\bar{q}^{(5)}\Big )_{(SF)_{\bar{D}^{(*)}}},
\en
where the superscript $(423(\rm{or}~143))$ indicates that quarks in $\Lambda_c$ are labeled by 4, 2, and 3
(or 1, 4, and 3), and $(1(2))$ and $(5)$ denotes that quarks in $\bar{D}^{(*)}$ are marked
by 1(or 2) and 5, respectively. The detailed forms of $\xi_{\rho,\lambda}$ (or $\chi_{\rho,\lambda}$) can also be
found in Ref.~\cite{close}.\\

\subsection{Numerical results} \par\noindent\par

The wave function of $P_c(4312)$ with a $\Sigma_c\bar{D}$ structure is solved in our previous bound state
calculation~\cite{Wang:2011rga}. In terms of this established wave function, the harmonic oscillator frequency $\omega$
is fixed as a universal value of $0.45$~GeV. Then, the width parameter of the different hadrons in the final state
can roughly be determined by the values of $\omega$ and different masses of quarks in corresponding hadrons, and
are tabulated in Table~{\ref{tab:widthparameter}}.
We first calculate the spin-flavor-color coefficients by using the spin-isospin wave functions in both initial and
final states shown in eqs.~({\ref{eq:spinisospin}}), ({\ref{eq:spinisospinB}}) and ({\ref{eq:spinisospinC}}) and
re-coupling coefficients. The obtained spin-flavor-color (SFC) coefficients are tabulated in Table~{\ref{tab:SFC}}.
\begin{table}
\caption{\label{tab:widthparameter} The size parameters of final
hadrons (in units of $\rm{fm}$) with the universal
$\omega=0.45~\rm{GeV}$ and the quark mass of $m_l=0.313~\rm{GeV}$
and $m_c=1.54~\rm{GeV}$.} \vspace{0.5cm}
\begin{center}
\begin{tabular}{|c|c|c||c|c|c|}\hline
&&&&&\\
$b_{c\bar{c}}$ &$b_{p1}$  &$b_{p2}$ &$b_{\bar{D}^*}$ &$b_{{\Lambda_c}1}$  &$b_{{\Lambda_c}2}$\\
&&&&&\\ \hline
&&&&&\\
$\sqrt{\frac{2}{\omega m_c}}$ &$\sqrt{\frac{2}{\omega m_l}}$
&$\sqrt{\frac{3}{2\omega m_l}}$ &$\sqrt{\frac{m_l+m_c}{\omega
m_lm_c}}$ &$\sqrt{\frac{2}{\omega m_l}}$
&$\sqrt{\frac{2m_l+m_c}{2\omega m_lm_c}}$\\
&&&&&\\ \hline
&&&&&\\
0.335  &0.744 &0.644 &0.577 &0.744 &0.441\\
&&&&&\\ \hline
\end{tabular}
\end{center}
\end{table}
Based on these width parameters, the final state wave functions written on the harmonic oscillator basis are entirely
fixed. Then, we can calculate the decays widths.
\begin{table}[htbp]
\caption{\label{tab:SFC} Color coefficient $F^c$ and spin-flavor
coefficient $F^{SF}$ for $P_c(4312)\to p+J/\psi\,(\eta_c)$ and
$P_c(4312)\to \Lambda_c+\bar{D}^{(*)}$, where $Dir.$,
$\hat{P}_{41}$, $\hat{P}_{42}$, and $\hat{P}_{43}$ stand for the
direct term, and permutations of (14), (24), and (34),
respectively.}
\begin{center}
\begin{tabular}{|c||c|c||c|c|||c|c||c|c||c|c||c|c|}\hline
&\multicolumn{4}{|c|||}{$P_c(4312)\to p+J/\psi\,(\eta_c)$}&
\multicolumn{8}{|c|}{$P_c(4312)\to \Lambda_c+\bar{D}^{(*)}$}\\
\cline{2-13}
&\multicolumn{2}{|c||}{$p+J/\psi$} &
\multicolumn{2}{|c|||}{$p+\eta_c$}
&\multicolumn{2}{|c||}{$\Lambda_c^{(423)}+\bar{D}^{(15)}$} &
\multicolumn{2}{|c||}{$\Lambda_c^{(423)}+\bar{D}^{*(15)}$}
&\multicolumn{2}{|c||}{$\Lambda_c^{(143)}+\bar{D}^{(25)}$} &
\multicolumn{2}{|c|}{$\Lambda_c^{(143)}+\bar{D}^{*(25)}$}\\
\cline{2-13}
 & $F^c$ & $F^{SF}$  & $F^c$ & $F^{SF}$  & $F^c$ & $F^{SF}$  &
$F^c$ & $F^{SF}$  & $F^c$ & $F^{SF}$  & $F^c$ & $F^{SF}$ \\ \hline
$Dir.$ & $+\frac{1}{3}$ & $+\frac{1}{2 \sqrt{3}}$ & $+\frac{1}{3}$  &
$\frac{1}{2}$& $+\frac{1}{3}$ & 0 & $+\frac{1}{3}$ & $-\frac{1}{2
\sqrt{3}}$ &
$+\frac{1}{3}$ & 0  & $+\frac{1}{3}$ & $-\frac{1}{2\sqrt{3}}$ \\
\hline
$P_{14}$ & $-\frac{1}{3}$ & $+\frac{1}{2 \sqrt{3}}$ & $-\frac{1}{3}$
& $\frac{1}{2}$ &  $+1$ & 0 & $+1$ & 0 &
$-\frac{1}{3}$ & 0 & $-\frac{1}{3}$ & $-\frac{1}{2\sqrt{3}}$  \\
\hline
$P_{24}$ & $-\frac{1}{3}$ & $+\frac{1}{2 \sqrt{3}}$ & $-\frac{1}{3}$
& $\frac{1}{2}$ & $-\frac{1}{3}$ & 0 &
$-\frac{1}{3}$ & $-\frac{1}{2\sqrt{3}}$ & $+1$ & 0 & $+1$ & 0  \\
\hline
$P_{34}$ & $+1$ & 0 & $+1$ & 0 &  $-\frac{1}{3}$ & 0 &
$-\frac{1}{3}$ & $+\frac{1}{2\sqrt{3}}$ & $-\frac{1}{3}$ & 0 &
$-\frac{1}{3}$ & $-\frac{1}{2\sqrt{3}}$  \\ \hline
\end{tabular}
\end{center}
\end{table}
Working out the spatial integrals together with the SFC coefficients
shown in Table~{\ref{tab:SFC}}, the partial decay widths can be
calculated and are tabulated in Table~{\ref{tab:decaywidth}}.\\
\begin{table}
\begin{center}
\caption{\label{tab:decaywidth} The calculated partial decay widths
(in units of $\rm{MeV}$) comparing to the experimental measurement.
The upper and lower uncertainties in the obtained
widths in the 1st, 2nd, 3rd, and 5th columns correspond to the
$+20(+30)\%$ and $-20(-30)\%$ uncertainties in the
averaged binding energy of $P_c(4312)$.} \vspace{0.5cm}
\begin{tabular}{|c|c|c|c||c|||c|}\hline
&&&&&\\
$\Gamma_{J/\psi p}$  &$\Gamma_{\eta_c p}$ &$\Gamma_{\bar{D}^*\Lambda_c}$
&$\Gamma_{\bar{D}\Lambda_c}$ &$\Gamma_{Total}$  &$\Gamma_{Expt.}$ \\
&&&&&\\ \hline
&&&&&\\
$0.0448^{+0.0197(+0.0309)}_{-0.0161(-0.0287)}$
&$0.0892^{+0.0392(+0.0615)}_{-0.0321(-0.0571)}$
&$8.36^{+3.68(+5.77)}_{-3.01(-5.35)}$  &0
&$8.49^{+3.74(+5.86)}_{-3.06(-5.43)}$  &$9.8\pm
2.7^{+3.5}_{-4.5}$\\
&&&&&\\ \hline
\end{tabular}
\end{center}
\end{table}
\vspace{0.5cm}

It should be addressed that the spatial overlap between the initial
and final states is model-dependent. However, the overlap in the
spin-isospin and in color spaces between those states comes from the
intrinsic properties of involved hadrons, including the quark
re-arrangement, and therefore, is model-independent. In
Tab.~{\ref{tab:SFC}}, it is shown that the spin-isospin factor in
the $p \eta_c $ channel, where the re-arrangement effect is
involved, is three times larger than that in the $p J/\psi $
channel. This is in agreement with the remarks in
Ref.~\cite{Voloshin:2019aut}. However, it is also shown in
Tab.~{\ref{tab:decaywidth}} that the ratio of the corresponding
partial decay widths is not exactly 3, but around 2, although the
same width parameter for both $J/\psi$ and $\eta_c$ in the final
state has been taken. This is because that the overlap in the
orbital space is closely related to the
outgoing momentum $k$, namely it is momentum-dependent.\\

Moreover, our calculation shows that the partial decay width of the
$(\bar{D}\Lambda_c)$ channel vanishes. This is due to the zero
overlap between the spin-isospin wave functions in the initial and
final states where the quark re-arrangement involved (see Tab. II).
This outcome consists with the results from most other models in the
molecular scenario on the hadronic degrees of freedom, where the
$t$-channel pion-exchange between the pseduo-scalar meson $\bar{D}$
and $\Lambda_c$ baryon does not exist. Finally, the
result in Tab.~\ref{tab:decaywidth} tells us that the decay channel
$\Lambda_c\bar{D}^*$ is dominant and the branching ratios of the
other two hidden charm decays are both
less than 1\%.\\

It is important to point out that the quark exchange effect comes
from the Pauli principle (see eq.(\ref{eq:antisym})). Although the
binding energy is small and the mass of $P_c(4312)$ is very close to
the threshold of $\bar{D}\Sigma_c$, we still see that the effect of
the quark exchange, from the last three terms on the right side of
eq.(\ref{eq:antisym}), plays a sizable role on the partial decays
width of $P_c(4312)$. For example, such effect reduces the width of
the $p\,J/\psi$ (or $p\,\eta_c$) decay from $0.163~\rm{MeV}$ (or
$0.313~\rm{MeV}$), which is solely contributed by the direct term
(the first term on the right side of eq.(\ref{eq:antisym})), to
$0.0448~\rm{MeV}$ (or $0.0892~\rm{MeV}$), which is a sum of all the
four terms on the right side of eq.(\ref{eq:antisym}). For the open
charm decays $\bar{D}^{(*)}\Lambda_c$, compared with the total
contribution given by the sum of all the terms, the quark exchange
effect also significantly reduces the width produced by the direct
term by a factor of 0.74. Therefore, it is a defect that such an
important re-arrangement effect can not be accounted for in the
calculations with the molecular scenario in the hadron
level~\cite{Lu:2016nnt,Yamaguchi:2016ote,Lin:2017mtz,Xiao:2019mst,Gutsche:2019mkg,He:2019rva,Wang:2019krd,Lin:2019qiv}.
So far, how to make a direct connection between the calculations on
the quark degrees of freedom and on the hadron degrees of freedom is
still an open question. Moreover, one may also notice that the
calculated branching ratios vary from tens of
percentages~\cite{Gutsche:2019mkg} to even 0.03\%~\cite{Lin:2017mtz}
in the literature. It is necessary to pin down which one is more
meaningful. Fortunately, our small $p\,J/\psi$ branching ratio is
compatible with the upper limit of 4.6\% given by the GlueX
Collaboration~\cite{Ali:2019lzf}, and the obtained total width of
$P_c$ of about $8.5$~MeV is compatible with the data of $9.8\pm
2.7^{+3.5}_{-4.5}$~MeV, although it is a rough estimate.\\

Since our results are model-dependent in spatial integrals, for giving a
reference to the effect of our approximation made on the matrix
elements of the interaction, we further calculate above widths by
taking $\pm 20\% (\pm 30\%)$ uncertainties in the
averaged binding energy of $P_c(4312)$ and also put them in
Tab.~{\ref{tab:decaywidth}}. The numbers in the superscript and
subscript denote the uncertainties in width with a
$+20\%(+30\%)$ and $-20\%(-30\%)$ uncertainties in the
averaged binding energy of $P_c(4312)$, respectively. It shows that
even with a large tolerance in $\overline{\cal H}_{Int}$, the
obtained widths with deviation are still compatible with the
experimental values with error. However, a sophisticated calculation
is urgent and necessary for finally identifying the structure of
$P_c$. Now, we expect that by measuring the total width, especially
the partial decay width of the open charm decays of $P_c(4312)$ such
as $P_c(4312) \to \bar{D}^*\Lambda_c$ or $P_c(4312) \to
\bar{D}\pi\Lambda_c$, more accurately, one may obtain
the criterion for branching ratios and the appropriateness for different models. \\

\section{Summary}\par\noindent\par

In this work, we show a calculation for the selected strong decays
of the newly observed $P_c(4312)$ with our chiral constituent quark
model which has been proved to have the predictive power because
most of existing data can be well-reproduced. The advantages of our
calculation are twofold. 1), we are working in the quark level and
the predicted  mass of $P_c(4312)$ in our previous work is
consistent with the observed data. 2), The wave function used in
this calculation was obtained in the same calculation where the
above mentioned mass is resulted without any additional parameters
or form factors introduced by hands. Therefore, we expect that in
comparison with other calculations, our model results are more
meaningful. The branching ratios of the $p\,J/\psi $ and $p\,\eta_c
$ decays in our model calculation are both less than 1\%, and the
latter is about 2 times larger than the former. In particular, the
$\bar{D}^* \Lambda_c$ decay mode dominates. We also find that the
quark exchange effect is sizeable. This is because that our
$P_c(4312)$ wave function is on the quark degrees of freedom, where
the effect of the quark re-arrangement can be considered explicitly.
Therefore, in order to understand the nature of $P_c$, it is
necessary to redo a calculation for the decays studied
in this work in a more accurate way and predict more properties of
$P_c$, such as the electromagnetic transition of $P_c(4312)$ as well
as three-body decays, like $P_c(4312)\rightarrow \bar{D}^*\pi\Lambda_c$, for further
experimental inspection. These studies are in progress.\\

\newpage
\section*{Acknowledgment}

\par\noindent\par
This work is supported by the National Natural Sciences Foundations of China under the  Grant Nos. 11521505, 11975245,
11975083, and 11635009, the Sino-German CRC 110 "Symmetries and the Emergence of Structure in QCD" project by NSFC under
the grant No.11621131001, the Key Research Program of Frontier Sciences, CAS, Grant No. Y7292610K1, and the IHEP
Innovation Fund under the grant No. Y4545190Y2. \\

\section*{Appendix}

The unitary transformation between coordinate sets
$(\vec{r}_1,\vec{r}_2,\vec{r}_3,\vec{r}_4,\vec{r}_5)$ and
$(\vec{R}_{AB},\vec{r}_{AB},\vec{\xi}_1,\vec{\xi}_2,\vec{\xi}_3)$ in
the initial state is \eq \left (\begin{array} {c}
\vec{r}_1\\
\vec{r}_2\\
\vec{r}_3\\
\vec{r}_4\\
\vec{r}_5
\end{array}\right )
=\left [\begin{array}{ccccc}
1 &\frac{m_{45}}{m_{1-5}} & -\frac{m_2}{m_{12}} &-\frac{m_3}{m_{123}} &0 \\
1 &\frac{m_{45}}{m_{1-5}} &+\frac{m_1}{m_{12}} &-\frac{m_3}{m_{123}} &0 \\
1 &\frac{m_{45}}{m_{1-5}} &0                              &+\frac{m_{12}}{m_{123}} &0 \\
1 &-\frac{m_{123}}{m_{1-5}} &0  &0 &-\frac{m_5}{m_{45}} \\
1 &-\frac{m_{123}}{m_{1-5}} &0  &0 &+\frac{m_{4}}{m_{45}}
\end{array}\right ]
\left (\begin{array} {c}
\vec{R}_{AB}\\
\vec{r}_{AB}\\
\vec{\xi}_1\\
\vec{\xi}_2\\
\vec{\xi}_3
\end{array}\right ),
\en
and the unitary transformation between coordinate sets
$(\vec{r}_1,\vec{r}_2,\vec{r}_3,\vec{r}_4,\vec{r}_5)$ and
$(\vec{R'}_{A'B'},\vec{r'}_{A'B'},\vec{\xi}'_1,\vec{\xi}'_2,\vec{\xi}'_3)$
in the final state is \eq \left (\begin{array} {c}
\vec{R}'_{A'B'}\\
\vec{r}'_{A'B'}\\
\vec{\xi}'_1\\
\vec{\xi}'_2\\
\vec{\xi}'_3
\end{array}\right )
=\left [\begin{array}{ccccc}
\frac{m_{1}}{m_{1-5}} &\frac{m_{2}}{m_{1-5}} & \frac{m_{4}}{m_{1-5}} &\frac{m_{3}}{m_{1-5}}&\frac{m_{5}}{m_{1-5}} \\
\frac{m_{1}}{m_{1-5}} &\frac{m_{2}}{m_{1-5}} & \frac{m_{4}}{m_{1-5}} &- \frac{m_{3}}{m_{1-5}}&- \frac{m_{5}}{m_{1-5}} \\
-1 & +1 & 0 & 0 & 0\\
- \frac{m_{1}}{m_{12}} &- \frac{m_{2}}{m_{12}} & +1 & 0 & 0 \\
0 & 0 & 0 & -1 & +1
\end{array}\right ]
\left (\begin{array} {c}
\vec{r}_1\\
\vec{r}_2\\
\vec{r}_4\\
\vec{r}_3\\
\vec{r}_5
\end{array}\right ).
\en

\end{document}